%% file: ms.tex
\documentclass{llncs}

\usepackage{frameit,amsmath,latexsym}
\usepackage{multirow}
\usepackage{array}
\newcolumntype{H}{>{\setbox0=\hbox\bgroup}c<{\egroup}@{}}

\usepackage{times}

\input{header}
\input{macros}

\newcommand{\conferencepaper}{0} 
\newcommand{\rep}[1]{\ifthenelse{\conferencepaper = 0}{#1}{}}
\newcommand{\conf}[1]{\ifthenelse{\conferencepaper = 0}{}{#1}}
\newcommand{\repconf}[2]{\ifthenelse{\conferencepaper = 0}{#1}{#2}}

\title{Concolic Testing Heap-Manipulating Programs \\ (Technical Report)}

\author{Long H. Pham\inst{1}\thanks{Corresponding author. Email: longph1989@gmail.com} \and Quang Loc Le\inst{2}  \and Quoc-Sang Phan\inst{3}  \and Jun Sun\inst{4}}
\institute{Singapore University of Technology and Design, Singapore \and School of Computing \& Digital Technologies, Teesside University, UK
\and {Synopsys, Inc., USA}
\and {Singapore Management University, Singapore}
}

\begin{document}

\maketitle
\begin{abstract}
\input{sections/abstract}
\end{abstract}

\section{Introduction}

\input{sections/intro}

\section{Approach at a Glance}\label{sec.mov.ex}

\input{sections/example}
\section{Specification-based Testing}\label{sec.background}

\input{sections/background}

\section{Concolic Execution}\label{sec.concolic}
\input{sections/details}

\section{Implementation and Experiments}\label{sec.exp}

\input{sections/experiment}

\section{Related Work}\label{sec.related}
\input{sections/relatedWork}

\section{Conclusion}\label{sec.concl}
\input{sections/conclusion}

\noindent \textbf{Acknowledgments.} This research is supported by MOE research grant MOE2016-T2-2-123.

\repconf{
\appendix

\subsection*{Appendix 1: Method \emph{repOK} for \emph{BinarySearchTree}}\label{app.A}
\begin{center}
\begin{minipage}[h]{0.85\linewidth}
\begin{lstlisting}[numbers=none]
public class BinarySearchTree {
	public boolean repOK(BinaryNode t) {
		return repOK(t, new Range());
	}
	boolean repOK(BinaryNode t, Range range) {
		if (t == null) return true;
		if (!range.inRange(t.element)) return false;
		boolean ret = true;
		ret = ret && repOK(t.left, range.setUpper(t.element));
		ret = ret && repOK(t.right, range.setLower(t.element));
		return ret;
	}
}
\end{lstlisting}
\end{minipage}
\end{center}

\subsection*{Appendix 2: Symbolic model to test input}\label{app.tounit}
\input{sections/tounit}

\subsection*{Appendix 3: Path condition transformation example}\label{app.C}
\input{sections/appC}

\subsection*{Appendix 4: Instructions to replicate the experiments}\label{app.D}
\input{sections/artifact}
}

\bibliographystyle{splncs04}
\bibliography{papers}

\end{document}

%% file: header.tex
\usepackage[colorinlistoftodos,prependcaption,textsize=tiny]{todonotes}
\usepackage{listings}

\usepackage[linesnumbered,ruled,vlined]{algorithm2e}

\newcommand{\tool}{CSF}

\newcommand{\lazy}{lazy initialization}

\newcommand{\sepnode}[3]{\ensuremath{#1{\mapsto}#2(#3)}}
\newcommand{\seppred}[2]{\ensuremath{#1(#2)}}
\newcommand{\emp}{\btt{emp}}
\def\sep{\ensuremath{*}}
\newcommand{\pto}{{\scriptsize\ensuremath{\mapsto}}}
\newcommand{\nil}{\btt{null}}

\newcommand{\code}[1]{{\small {\ensuremath{\tt #1}}}}
\newcommand{\form}[1]{\ensuremath{#1}}
\newcommand{\btt}[1]{{\ensuremath{\tt #1}}}

\def\bool{\code{bool}}
\def\int{\code{int}}

\newcommand{\setvars}[1]{\ensuremath{\bar{#1}}}
\newcommand{\constr}{\ensuremath{\Phi}}
\def\D{\Delta}
\newcommand{\heap}{\ensuremath{\kappa}}
\newcommand{\atom}{\alpha}
\newcommand{\sepnodeF}[3]{\ensuremath{{#1}{\pto}#2(#3)}}
\newcommand{\seppredF}[2]{\ensuremath{#1(#2)}}
\newcommand{\hide}[1]{}
\def\a{a}
\def\true{\code{true}}
\def\false{\code{false}}
\newcommand{\myit}[1]{\textit{#1}}

\newcommand{\pres}{\ensuremath{\Gamma}}

\newcommand{\sstack}{\ensuremath{s}}

\newcommand{\pure}{\ensuremath{\pi}}
\newcommand{\subst}[2]{\ensuremath{[#1 {/} #2]}}
\newcommand{\self}{\btt{root}}
\newcommand{\yields}{\leadsto}
\newcommand{\myif}[3]{\code{if}~#1~\code{then}~#2~\code{else}~#3}

\newcommand{\rulen}[1]{\ensuremath{{\bf \scriptstyle [#1]}}}
\newcommand{\hlr}[3]{\ensuremath{\rulen{#1}\frac{
#2
}{#3}}}
\newcommand{\hquad}[2]{\ensuremath{#1 ~{\yields}~ #2}}

\def\solver{\form{\code{S2SAT_{SL}}}}

\newcommand{\sectx}[5]{\ensuremath{\langle#1,#3,#2,#4,#5\rangle}}

\newcommand{\evalse}[3]{\ensuremath{#1 ~{\vdash}~ #2 \Downarrow#3}}

\def\genSpec{\form{\code{genFromSpec}}}
\def\unfold{\code{unfold}}

\def\tounit{\form{\code{toUnitTest}}}

\newcommand{\norm}[1]{\ensuremath{{#1}}}

%% file: macros.tex
\graphicspath{{figs/}}
\definecolor{mygreen}{rgb}{0,0.6,0}
\definecolor{mygray}{rgb}{0.5,0.5,0.5}
\definecolor{mymauve}{rgb}{0.58,0,0.82}

%% file: sections/abstract.tex
Concolic testing is a test generation technique which works effectively by integrating random testing generation and symbolic execution. Existing concolic testing engines focus on  numeric programs. Heap-manipulating programs make extensive use of complex heap objects like trees and lists. Testing such programs is challenging due to multiple reasons. Firstly, test inputs for such programs are required to satisfy non-trivial constraints which must be specified precisely. Secondly, precisely encoding and solving path conditions in such programs are challenging and often expensive. In this work, we propose the first concolic testing engine called CSF for heap-manipulating programs based on separation logic.
CSF effectively combines specification-based testing and concolic execution for test input generation. It is evaluated on a set of challenging heap-manipulating programs. The results show that CSF generates valid test inputs with high coverage efficiently. Furthermore, we show that CSF can be potentially used in combination with precondition inference tools to reduce the user effort.

%% file: sections/intro.tex
Unit testing is essential during the software development process. 
To automate unit testing effectively, we are required to generate \emph{valid} test inputs which exercise program behaviors \emph{comprehensively} and \emph{efficiently}. Many techniques for automating unit testing have been proposed, including random testing~\cite{Godefroid:2005:DDA:1065010.1065036} and symbolic execution~\cite{Visser:ISSTA:2004}. A recent development is the concolic testing technique~\cite{tacas2016-ldghikrr,Sen2005}. Concolic testing works by integrating random testing and symbolic execution to overcome their respective limitations~\cite{wang2018towards}. It has been shown that concolic testing often works effectively~\cite{DBLP:conf/uss/Yun0XJK18}.

Existing concolic testing engines focus on numeric programs, i.e., programs which take numeric type variables as inputs.
In contrast, \emph{heap-manipulating} programs make extensive use of heap objects and their inputs are often dynamically allocated data structures.
Test input generation for heap-manipulating programs is hard for two reasons. Firstly, the test inputs are often heap objects with complex structures and strict requirements over their shapes and sizes.
 Secondly, the inputs have unbounded domains.
Ideally, test generation for heap-manipulating programs must satisfy three requirements.
\begin{enumerate}
    \item \emph{(Validity)} It must generate valid test inputs.
    \item \emph{(Comprehensiveness)} It must exercise program behaviors comprehensively, e.g., maximizing certain code coverage.
    \item \emph{(Efficiency)} It must be efficient. 
\end{enumerate}
Existing approaches often overlook one or more of the requirements. The state-of-the-art approaches are based on classical symbolic execution~\cite{King:1976:SEP:360248.360252} with \lazy~\cite{Visser:ISSTA:2004}. To achieve comprehensiveness and efficiency, lazy initialization postpones the initialization of reference type symbolic variables and fields until they are accessed.
However, \lazy~has
 limited support to capture constraints on the shapes of the input data structures. As a result, invalid test inputs are generated, which are not only wasteful but also lead to the exploration of infeasible program paths.
Furthermore, because the values of un-accessed fields are not initialized, the generated test inputs need to be further concretized.
Subsequent works on improving \lazy~~\cite{Deng:ASE:2006,Deng:2007:TCS:1306879.1307404,Hillery:2016:EHS:2963187.2963200,Visser:ISSTA:2004} share the same aforementioned problems. To address the validity requirement, Braione \emph{et al.}~\cite{Braione:2015:SEP} introduced a logic called HEX as a specification language for the input data structures. However, HEX has limited expressiveness and thus cannot describe many data structures (unless using additional user-provided methods called $triggers$).

Inspired by the recent success of concolic execution (e.g.,~\cite{PySymEmu,DBLP:conf/ndss/StephensGSDWCSK16}),
we aim to develop a concolic execution engine for heap-manipulating programs. Developing a concolic execution engine which achieves validity, comprehensiveness and efficiency
is however highly non-trivial.
For validity, we need a specification language which is expressive enough to capture constraints over the shapes and sizes of heap objects. We thus adopt a recently proposed fragment of separation logic which is shown to be expressive and decidable~\cite{DBLP:conf/cav/LeT0C17}. For comprehensiveness and efficiency, we propose a novel concolic testing strategy which combines specification-based testing and concolic execution. That is, we first generate test inputs according to the specification in a black-box manner and then apply concolic execution to cover those uncovered program parts. 

In summary, we make the following contributions. Firstly, we propose a concolic execution engine for heap-manipulation programs based on separation logic. 
Secondly, we combine specification-based testing with concolic execution in order to reduce the cost of constraint solving.
Thirdly, we implement the proposal in a tool called Concolic StarFinder (CSF) and evaluate it in multiple experiments.

The rest of this paper is organized as follows.
Section \ref{sec.mov.ex} illustrates
our approach through an example.
Section \ref{sec.background} describes 
our specification language
and specification-based test input generation. 
Next, we present our concolic execution engine
in Section \ref{sec.concolic}. We show the implementation
and experiments in Section \ref{sec.exp}.
Section \ref{sec.related} discusses related works
and finally Section \ref{sec.concl}
concludes.

%% file: sections/example.tex
We illustrate our approach using method $remove$ in class $BinarySearchTree$ from the SIR repository~\cite{sir}. The method is shown in Fig.~\ref{src:remove}. It checks if a binary search tree object contains a node with a specific value and, if so, removes the node. To test the method, we must generate two inputs, i.e., a \emph{valid} binary search tree object $t$ and an integer $x$, and then execute \emph{t.remove(x)}. Note that a valid binary search tree object must satisfy strict requirements. Firstly, all \emph{BinaryNode} objects must be structured in a binary tree shape. Secondly, for any \emph{BinaryNode} object in the tree, its \emph{element} value must be greater than all the \emph{element} values of its \emph{left} sub-tree and less than those of the \emph{right} sub-tree. One way to define valid binary search tree objects is through programming a $repOK$ method~\cite{Boyapati:2002:KAT:566172.566191,Visser:ISSTA:2004}\repconf{, as shown in App. 1.}{.}
\begin{figure}[t]
\centering
\begin{minipage}[t]{0.85\linewidth}
\begin{lstlisting}
public class BinarySearchTree {
	public BinaryNode root;	
	public void remove(int x) {
		root = remove(x, root);
	}	
	private BinaryNode remove(int x, BinaryNode t) {
		if (t == null) return t; @\label{t1}@
		if (x < t.element)
			t.left = remove(x, t.left); @\label{t2}@
		else if (x > t.element) 
			t.right = remove(x, t.right);
		else if (t.left != null && t.right != null){
			t.element = findMin(t.right).element;
			t.right = remove(t.element, t.right);
		} else 
			t = (t.left != null) ? t.left : t.right;
		return t;
	}			
	private BinaryNode findMin(BinaryNode t) {
		if (t == null) return null;
		else if (t.left == null) return t; @\label{x1}@
		return findMin(t.left);
	}
}

public class BinaryNode {
	int element; BinaryNode left; BinaryNode right;	
}
\end{lstlisting}
\end{minipage}
\caption{Sample program}
\label{src:remove}
\end{figure}


If a $repOK$ method is provided, we can use the black-box enumeration (BBE) approach~\cite{Visser:ISSTA:2004} to generate test inputs. BBE performs symbolic execution with lazy initialization on the $repOK$ method. Although BBE can generate valid test inputs, it also generates many invalid ones, 
e.g., the generated input is a cyclic graph instead of a tree\footnote{When BBE runs, we count the structures that the $repOK$ method returns \true~as valid ones, and the structures
that the $repOK$ method returns \false~as invalid ones.}.
In our experiment with BBE for this method, a total of 225 test inputs are generated and only 9 of them are valid. Moreover, because BBE generates test inputs based on the $repOK$ method only, it may not generate a high coverage test suite.

One way to obtain a high coverage test suite is to use the white-box enumeration approach~\cite{Visser:ISSTA:2004}. First, white-box enumeration performs symbolic execution on the method under test to create some partially initialized data structures. Then, these data structures are used as initial inputs to perform symbolic execution with the $repOK$ method. However, because the approach still uses lazy initialization, many invalid test inputs may be generated.
Moreover, white-box enumeration requires the availability of a conservative $repOK$ method in the first step, which is not easy to derive.
Another approach is to use the HEX logic~\cite{Braione:2016:JSE:2950290.2983940} as a language to specify valid data structures. During lazy initialization, the exploration is pruned when the heap configuration violates the specification. However, HEX has limited expressiveness, e.g., HEX cannot capture the property that the nodes in the binary search tree are sorted due to the lack of arithmetic constraints.



In comparison, our approach works as follows. We use separation logic to 
define a
predicate \seppred{\code{bst}}{\self,minE,maxE}, which specifies valid binary search trees
where \form{\self} is the root of the tree and \form{minE} (resp. \form{maxE})
is the minimum (resp. maximum) bound of the $element$ values of the tree. We refer the readers to Section~\ref{sec.background} for details of the definition. The precondition of method $remove$ is then specified as \form{\seppred{\code{bst}}{this\_root,minE,maxE}}. With the specification, we first apply specification-based testing based on the precondition in a black-box manner. That is, we generate the test inputs according to the precondition using a constraint solver without exploring the method body. After this step, we generate 22 test inputs and they cover 14 over 15 feasible branches of the method $remove$ (including auxiliary method $findMin$). The only branch which is not covered is the $else$ branch at line~\ref{x1}. We then perform concolic execution with the generated test inputs to identify a feasible path which leads to the uncovered branch. After solving that path condition, we obtain the test inputs for 100\% branch coverage.

%% file: sections/background.tex
Our approach takes as input a heap-manipulating program which has a precondition specified using a language recently developed in~\cite{Chin:SCP:2012,DBLP:conf/cav/LeT0C17}. In the following, we introduce the language and present the first step of our approach, i.e., specification-based testing based on the provided precondition. \\

\input{sections/background-sl}


\input{sections/blackbox-gen}

%% file: sections/background-sl.tex
\input{sections/fig-sl-syn}

\noindent \textbf{Specification Language}
The language we adopt supports separation logic, inductive predicates and arithmetical constraints, 
which is 
expressive to specify many data structures~\cite{Chin:SCP:2012,DBLP:conf/cav/LeT0C17}.
Its syntax is shown in Fig.~\ref{prm.sl.fig}. In general, the precondition 
is a disjunction of one or more symbolic heaps.
A symbolic heap is an existentially quantified conjunction of a heap formula \heap~and a pure
formula \pure. While a pure
formula is a constraint in the form
of the first-order logic, the heap formula
is a  conjunction of heap predicates which are
connected by separating operation {\sep}.
A heap predicate may be the empty predicate \form{\emp}, a
points-to predicate \form{\sepnodeF{x}{c}{\setvars{v}}}
or an inductive predicate
 \form{\seppredF{\code{P}}{\setvars{v}}}.
Reference types
are annotated by the keyword \form{\code{data}}. Variables
may have type
\form{\tau} as
boolean \form{\code{bool}} or 32-bit integer \form{\code{int}} or user-defined reference type \form{\code{c}}.

Inductive predicates are supplied by the users with the keyword \form{\code{pred}}. They are used to specify constraints on recursively defined data structures like linked lists or trees. Inductive predicates are defined
in the same language.
For instance, the inductive predicate \seppred{\code{bst}}{\self,minE,maxE} introduced in Section~\ref{sec.mov.ex} is defined as follows
\[
\small
\begin{array}{l}
\seppred{\code{pred~bst}}{\self,minE,maxE} ~{\equiv}~ (\emp \wedge \self = \nil) \\
\qquad \vee~ (\exists elt, l, r.~ \sepnodeF{\self}{BinaryNode}{elt,l,r} ~{\sep} \\
 \qquad \qquad \seppred{\code{bst}}{l,minE,elt} \sep \seppred{\code{bst}}{r,elt,maxE}~{\wedge}~ 
 minE < elt \wedge maxE > elt)
\end{array}
\]
, where \form{\self} is the root of the tree and \form{minE} (resp. \form{maxE})
is the minimum (resp. maximum) bound of the $element$ values of the tree. \hide{The predicate \form{\code{bst}} is a disjunction of two cases.
In the base case, \form{\emp} means empty heap and $\self$ equals to $\nil$. In this case, the values of $minE$ and $maxE$ are irrelevant.
In the inductive case, \form{\sepnodeF{\self}{BinaryNode}{elt,l,r}}
is a points-to predicate. It states that the $\self$ points to a $BinaryNode$ with three fields, e.g., $elt$, $l$, and $r$.
The value of $elt$ is greater than $minE$ and less than $maxE$.
Two fields $l$ and $r$ represent left and right subtree of $\self$ respectively and they are
constrained by $\form{\code{bst}}$ predicate.
Note that in the $\form{\code{bst}}$ constraint for field $l$, we replace $maxE$ with $elt$
to represent that the values of the left subtree should be less than that of the $\self$.
Similarly, we replace $minE$ with $elt$ in the $\form{\code{bst}}$ constraint for field $r$. Afterwards,}
 Using this definition with $this\_root$ as symbolic value for
field $root$ in class $BinarySearchTree$, the precondition of method $remove$ in the preceding section is then specified as
$\form{\seppred{\code{bst}}{this\_root,minE,maxE}}$. \\

%
%
%

%% file: sections/fig-sl-syn.tex
 \begin{figure}[t]
 \small
 \centering
 \[
\begin{array}{llr}
\text{Formula}            & \constr ::= \D ~|~ \constr_1 ~{\vee}~ \constr_2 & \\
\text{Symbolic heap}      & \D ::= \exists \bar{v}{.}~(\heap\wedge\pure) & \\
\text{Spatial formula}    & \heap ::= \emp ~|~ \sepnodeF{x}{c}{\setvars{v}} ~|~ \seppredF{\code{P}}{\setvars{v}} ~|~\heap_1\sep\heap_2 & \\
\text{Pure formula}       & \pure ::=  \true \mid \atom \mid {\neg} \pure \mid 
                           \pure_1 \wedge \pure_2 & \atom ::=   \a_1 = \a_2 \mid \a_1 \leq \a_2 \\
\text{}         &  
                           a ::= \nil \mid \!k \mid v \mid k {\times}\a \mid \a_1 \!+\! \a_2 \mid - \a & \\
\text{Data structure}     & \textit{Node} ::= \code{data}~ c_i \{\tau_1 ~ f_{i_1}; ...; \tau_j~f_{i_j} \} 
                         & \tau ::= \code{bool} \mid \code{int} \mid c \\
\text{Predicate definition}~~     & \textit{Pred} ::= \seppred{\code{pred~P_i}}{\setvars{v}_i}  \equiv \constr_i &
\end{array}
\]
\caption{Specification language, where \form{k} is a 32-bit integer constant, \form{\setvars{v}} is a sequence of variables}\label{prm.sl.fig}
\end{figure}

%% file: sections/blackbox-gen.tex
\input{sections/algo-genspec}
\noindent \textbf{Specification-based Testing} If we follow existing concolic testing strategies~\cite{Godefroid:2005:DDA:1065010.1065036}, we would first generate random test inputs 
 before applying concolic execution. However, it is unlikely that randomly generated heap objects are valid due to the strict precondition.
Thus, we apply
specification-based testing to generate test inputs based on the user-provided precondition instead.

The details are shown in Algorithm~\ref{algo.genSpec}.
The inputs are a set of formulae $\pres$ and a bound on $n$. The initial value of $\pres$ contains only the precondition of the program under test.
The output is a set of test inputs which are both \emph{valid} and \emph{fully initialized}.
Algorithm~\ref{algo.genSpec} has two phases. 
%

In the first phase, from line 8 to 12,
 procedure \form{\unfold} is applied to each symbolic heap $\D$ in {\pres} (at line 11) to return a set of unfolded formulae. Recall that a symbolic heap is 
 a conjunction of a heap constraint \heap~and a pure constraint \pure. If the heap constraint \heap~contains no inductive predicates (i.e., it is a base formula), \heap~is returned as it is. Otherwise, each inductive predicate \seppredF{\code{P}_i}{\setvars{t}_i} in \heap~is unfolded using its definition. Note that the definition of \seppredF{\code{P}_i}{\setvars{t}_i}~is
 a disjunction of multiple base cases and inductive cases. During unfolding, \heap~is split into a set of formulae, one for each disjunct in the definition of every inductive predicate \seppredF{\code{P}_i}{\setvars{t}_i} in \heap.
The process ends when $n$ reaches 0.

Procedure \form{\unfold} is formalized as follows.
Given an inductively predicate definition \form{\seppred{\code{pred~P}_i}{\setvars{v}_i}  \equiv \constr_i}
and a formula constituted with
 this predicate, e.g., \form{\D_i \sep \seppredF{\code{P}_i}{\setvars{t}_i}}, 
 \form{\unfold} proceeds in two steps.
First, it replaces the occurrences of the inductive predicate
with its definition as:
\form{\unfold(\D_i \sep \seppredF{\code{P}_i}{\setvars{t}_i}, \seppredF{\code{P}_i}{\setvars{t}_i}) \equiv \D_i * (\constr_i\subst{\setvars{t}_i}{\setvars{v}_i})}.
After that, it applies the following axioms to
normalizes the formula into
the grammar in Fig. \ref{prm.sl.fig}:
\[
\small
\begin{array}{lcl}
\norm{(\heap_1 \wedge \pure_1) \sep (\heap_2 \wedge \pure_2)}
& {\equiv}& \norm{(\heap_1 \sep \heap_2) \wedge (\pure_1 \wedge \pure_2)} \\
\norm{({\exists}  \setvars{w}.~\D_1) \sep ({\exists}  \setvars{v}.~ \D_2)}
&{\equiv}&   \norm{{\exists}\setvars{w},\setvars{v}'.~ (\D_1 \sep \D_2\subst{\setvars{v}'}{\setvars{v}})} \\
\end{array}
\]
The correctness of these axioms
could be found in~\cite{Ishtiaq:2001:BAL:360204.375719,Reynolds:LICS02}.
We then use \form{\unfold(\D) \equiv \bigcup_{i=1}^n \unfold(\D, \seppredF{\code{P}_i}{\setvars{t}_i}), \seppredF{\code{P}_i}{\setvars{t}_i} \in \D}.
For example, given the above-specified precondition for method $remove$
, we obtain 6 formulae shown in Fig.~\ref{unfoldings} after unfolding twice. \\
\begin{figure}[t]
{\small \[
\begin{array}{ll}
1. & \emp \wedge this\_root = \nil \\
2. & \exists elt, l, r.~ \sepnodeF{this\_root}{BinaryNode}{elt,l,r} \sep \seppred{\code{bst}}{l,minE,elt} \sep \seppred{\code{bst}}{r,elt,maxE} ~\wedge\\
   & \qquad minE < elt \wedge maxE > elt \\
3. &
\exists elt, l, r.~ \sepnodeF{this\_root}{BinaryNode}{elt,l,r} \sep \seppred{\code{bst}}{r,elt,maxE} \wedge l = \nil ~\wedge\\
   & \qquad minE < elt \wedge maxE > elt \\
4. & \exists elt, l, r, elt1, l1, r1.~ \sepnodeF{this\_root}{BinaryNode}{elt,l,r} \sep \sepnodeF{l}{BinaryNode}{elt1,l1,r1} ~\sep\\
   & \qquad \seppred {\code{bst}}{r,elt,maxE} \sep \seppred{\code{bst}}{l1,minE,elt1} \sep \seppred{\code{bst}}{r1,elt1,elt} ~\wedge\\
   & \qquad minE < elt \wedge maxE > elt \wedge minE < elt1 \wedge elt > elt1 \\
5. & \exists elt, l, r.~ \sepnodeF{this\_root}{BinaryNode}{elt,l,r} \sep \seppred{\code{bst}}{l,minE,elt} \wedge r = \nil ~\wedge\\
   & \qquad minE < elt \wedge maxE > elt \\
6. & \exists elt, l, r, elt2, l2, r2.~ \sepnodeF{this\_root}{BinaryNode}{elt,l,r} \sep \sepnodeF{r}{BinaryNode}{elt2,l2,r2} ~\sep\\
   & \qquad \seppred {\code{bst}}{l,minE,elt} \sep \seppred{\code{bst}}{l2,elt,elt2} \sep \seppred{\code{bst}}{r2,elt2,maxE} ~\wedge\\
   & \qquad minE < elt \wedge maxE > elt \wedge elt < elt2 \wedge maxE > elt2 
\end{array}
\]}
\caption{Unfoldings}
\label{unfoldings}
\end{figure}
%


\noindent \textbf{Unit Test Generation}
After unfolding, {\pres} contains a set of formulae, each of which satisfies the precondition. In the second phase, at lines 1-7, these formulae are transformed into test inputs. 
First, we check the satisfiability of each formula
using a satisfiability solver {\solver}~\cite{Le:CAV:2016,DBLP:conf/cav/LeT0C17} at line 4. 
The result of the solver is a pair (\form{\code{r}}, \form{\code{model}}) where \form{\code{r}} is a {\em decision} of satisfiability
and \form{\code{model}} is a symbolic model which serves as the evidence of the satisfiability.
Intuitively, a symbolic model is a base formula where every variable is assigned a {\em symbolic} value.
Formally, a symbolic model is a quantifier-free base formula \form{\D_m}
where \form{\D_m} is satisfiable and
for each variable \form{v} in \form{\D_m}, if \form{v} has a reference type, \form{\D_m} contains \form{\sepnodeF{v}{c}{...}}, or \form{v=v'}, or \form{v=\nil};
otherwise, \form{\D_m} contains \form{v=k} with $k$ is either a boolean or 32-bit integer constant.
%

At line 6, the symbolic model is transformed into a test input using procedure {\tounit},
\repconf{which we present in details in App. 2.}
{which initializes the variables according to the symbolic model
(e.g., for each points-to predicate \form{\sepnodeF{v}{c}{...}},
a new object of type $c$ is created and assigned to $v$).}
Fig.~\ref{tests} shows two test inputs generated for the example shown in Fig.~\ref{src:remove}. These two test inputs correspond to the first two formulae shown in Fig.~\ref{unfoldings} (where $x$ is assigned the default value 0).
\begin{figure}[t]
\centering
\begin{minipage}[h]{0.85\linewidth}
\begin{lstlisting}[numbers=none]
public void test_remove1() throws Exception {
	BinarySearchTree obj = new BinarySearchTree();
	obj.root = null; int x = 0;
	obj.remove(x);
}

public void test_remove2() throws Exception {
	BinarySearchTree obj = new BinarySearchTree();
	obj.root = new BinaryNode();
	BinaryNode left_2 = null; BinaryNode right_3 = null;
	int element_1 = 0; int x = 0; obj.root.element = element_1; 
	obj.root.left = left_2; obj.root.right = right_3;
	obj.remove(x);
}
\end{lstlisting}
\end{minipage}
\caption{Two test inputs}
\label{tests}
\end{figure}

The correctness of the algorithm, i.e.,
each generated test input is a valid one, 
is straightforward as
 each symbolic model obtained from the unfolding satisfies the original precondition, since each one is
an under-approximation of a \form{\D}~in \pres. 


%% file: sections/algo-genspec.tex
\begin{algorithm}[t]
	\small
\If{$n = 0$} {
  $tests \leftarrow \emptyset$\\
  \ForEach{$\D \in \pres$}{
    $\code{r},\code{model} \leftarrow \code{sat}(\D)$ \\
    \If{$\code{r} = \code{SAT}$} {
      $tests \leftarrow tests \cup \tounit(\code{model})$\\
    }
  }
  \Return ~$tests$\\
} \Else {
  $\pres' \leftarrow \emptyset$\\
  \ForEach{ $\D \in \pres$} {
    $\pres' \leftarrow \pres' \cup \unfold(\D)$\\
  }
  \Return $\genSpec(\pres', n - 1)$\\
}
\caption{$\genSpec(\pres, n)$}\label{algo.genSpec}
\end{algorithm}

%% file: sections/details.tex
Specification-based testing allows us to generate test inputs which cover some parts of the program. Some program paths however are unlikely to be covered with such test inputs without exploring the program code~\cite{wang2018towards}. Thus, the second step of our approach is to apply concolic execution to cover the remaining parts of the program.

We take a program, a set of test inputs and a constraint tree as inputs. The constraint tree allows us to keep track of both explored nodes and unexplored nodes. Informally, the concolic execution engine executes the test inputs, expands the tree and then generates new test inputs to cover the unexplored parts of the tree. This process stops when there are no unexplored nodes in the tree or it times out. 

\input{sections/prog-fig}

For simplicity, we present our concolic engine based on a general core intermediate language.
 The syntax of the language is shown in Fig.~\ref{fig.syntax}, which covers common programming language features.
A program in our core language
 includes several data structures and statements.
Our language supports boolean and 32-bit integer as primitive types.
Program statements include assignment, memory store, goto, assertion,
conditional goto, memory allocation, and memory deallocation.
Expressions are side-effect free
and consist of typical non-heap expressions and memory load.
We use \form{\code{op_{b}}} to represent binary operators, e.g.,
addition and subtraction, and \form{\code{op_{u}}} to represent
unary operators, e.g., logical negation.
 \form{k} is either a boolean or 32-bit integer constant.

We assume the program is in the form of static single assignments (SSA)
and omit the type-checking semantics of our language (i.e., we assume programs are
well-typed in the standard way).
Note that our prototype implementation is for Java bytecode, which in general can be translated to the core language
(with unsupported Java language features are abstracted during the translation).
The core language is easily extended to interprocedural scenario with method calls.\\

\noindent \textbf{Execution Engine}
Our concolic execution engine incrementally grows the constraint tree. Formally, the constraint tree 
is a pair \form{(V, E)} where
 $V$ is a finite set of nodes and 
 $E$ is a set of labeled and directed edges \form{(v, l, v')} where \form{v'} is a child of \form{v}.
Having edge \form{(v, l, v')} means that we can transit from \form{v} to \form{v'} via an execution rule \form{l}.
Each node in the tree is a concolic state in the form of a 6-tuple \form{\langle\Sigma, \D, \sstack, pc,flag\rangle\iota}
where \form{\Sigma} is the list of program statements;
\form{\D} is the symbolic state (a.k.a. the path condition);
\form{\sstack} is the current valuation of the program variables (i.e., the stack);
\form{pc} is the program counter; \form{flag} is a flag indicating whether
the current node has been explored or not and
\form{\iota} is the current statement.
Note that \form{\Sigma} and \form{\sstack} are mapping functions, i.e., \form{\Sigma} maps a number to a statement,
and \form{\sstack} maps a variable to its value.

Initially, the constraint tree has only one node \form{\langle\Sigma, \form{\code{pre}}, \emptyset, 0, \form{\code{true}}\rangle\iota_0} where \form{\emptyset} denotes an empty mapping function and \form{\iota_0} is the initial statement.
Note that the initial symbolic state is the precondition.
We start with executing the program concretely, with some initial test inputs (at least one),
and build the constraint tree along the way. The initial test inputs may come from specification-based testing
or be provided by the users.
Before each execution, $s$ is initialized with values according to the test input.
In the execution process, given a node,
our engine systematically identifies an applicable rule (based on the current statement) to generate one or more new nodes. 
If no rule matches (e.g., accessing a dangling pointer),
the execution halts. Note that some of the generated nodes are marked explored whereas some are marked unexplored (depending on the outcome of the concrete execution).

\input{sections/rules}

After executing all initial test inputs, the engine searches for unexplored nodes in the tree. If there is one such node with symbolic state \form{\D}, the engine solves
\form{\D}
using a solver~\cite{Le:CAV:2016,DBLP:conf/cav/LeT0C17}.
If
\form{\D} is satisfiable, the unexplored path is feasible and the symbolic model generated by the solver is transformed into a new test input (as shown in the Sect. \ref{sec.background}). The new test input is then executed and the constraint tree is expanded accordingly. If
\form{\D} is unsatisfiable, the node is pruned from the tree. 
This process is repeated until there are no more unexplored nodes or it times out.



The growing of the tree is governed by the execution rules, which effectively defines the semantics of our core language. The detailed execution rules are presented in Fig.~\ref{fig.rules}. 
One or more rules may be defined for each kind of statements in our core language.
Each rule, applied  based on syntactic pattern-matching, is of the following form.
\[
\begin{array}{c}
\text{conditions} \\
\hline
\text{current\_state}~\yields~\text{end\_state}_1, ... , \text{end\_state}_n
\end{array}
\]
Intuitively, if the conditions above the line is satisfied, a node matching the current\_state generates multiple children nodes.

In the following, we explain some of the rules in detail.
In the rule \form{\rulen{C-ASSIGN}} which assigns
the value evaluated from expression \form{e} to
variable \form{v}, for the concrete
state our system first evaluates the value of \form{e}
based on the concrete state \form{\sstack}
prior to updating the state of \form{v} with the new value.
For the symbolic state, it substitutes the current value
of \form{v} to a fresh symbol \form{v'}
prior to conjoining the constraint for the latest value
of \form{v}.
In the rule \form{\rulen{C-NEW}} which assigns new allocated
object to variable \form{v}, for the concrete
state our system updates the stack with an assignment
of the variable to a fresh location.
For the symbolic state, it substitutes the current value
of \form{v} to a fresh symbol \form{v'}
prior to spatially conjoining the points-to predicate
 for the latest value
of \form{v}.


In the rule \form{\rulen{C-LOAD}} (resp. \form{\rulen{C-STORE}}) which
reads from (resp. writes into) the field \form{f_i} of an object
\form{v}, in the concrete state we implicitly
assume that the corresponding variable of the field is \form{l.f_i}
where \form{l} is the concrete address of \form{v}
and proceed accordingly.
For the symbolic states, checking whether a variable has been allocated
before accessed is much more complicated as the path condition
(with the precondition) may include occurrences of inductive predicates
(which represent unbounded heaps), so our system keeps
the constraints with the field-access form (i.e., \form{v.f_i})
and field-assign form (i.e., $v.f_i := e$)
and will eliminate them before sending these formulae to the solver.

In the rule \form{\rulen{C-TCOND}},
 two new nodes denoting the \form{\code{then}}
 branch and the \form{\code{else}} branch
 of the condition are added into the tree with the current node is their parent.
The symbolic states
(path conditions) of both nodes are updated accordingly (\form{\D_1}
and \form{\D_2}).
The concrete state \form{\sstack} helps to identify that
the execution is going to follow the \form{\code{then}} branch
 and marks this branch as explored. The remaining node is marked as unexplored. The rule \form{\rulen{C-FCOND}} is interpreted similarly.

For example, Fig.~\ref{fig:tree} show the constraint trees constructed during the concolic execution of the example in Fig.~\ref{src:remove} with two initial test inputs in Fig.~\ref{tests}. The input of the first test case is an empty tree. The condition of the \form{\code{if-statement}} at line~\ref{t1} evaluates to \form{\code{true}}, satisfying the rule \form{\rulen{C-TCOND}}. The constraint tree in Fig.~\ref{fig:tree}(a) is constructed. The input of the second test case is a tree with one node and $x$ is 0. Thus the node is to be removed as its $element$ is $0$.
The rule \form{\rulen{C-FCOND}} is applied, which results in the tree in Fig.~\ref{fig:tree}(b). The condition $x < t.element$ is then used to generate a new test input with $x = 0$ and $t.element = 1$. Executing this new test input triggers the rule \form{\rulen{C-TCOND}} at line~\ref{t2}, and updates the constraint tree as in Fig.\ref{fig:tree}(c). \\

\input{sections/tree.tex}

\noindent \textbf{Path Condition Transformation} Note that the path conditions generated according to the execution rules may contain
field-access
and field-assign expressions
which are beyond the syntax in Fig.~\ref{prm.sl.fig} and the support of
the solver~\cite{Le:CAV:2016,DBLP:conf/cav/LeT0C17}. Thus, 
these
expressions 
 need to be eliminated.
The details of the transformation are presented in the Algorithm~\ref{algo.field.acc}.
The input of the algorithm is
a path condition which may contain field-access and field-assign expressions.
The output are multiple path conditions, i.e., a disjunction of path conditions, without
field-access and field-assign expressions.

The algorithm begins by
recording all symbolic values for all fields of points-to predicates (lines 1-3).
Then it considers each conjunct, which in form of a binary expression with left-hand side
and right-hand side, in the path condition (line 4).
In general, the field-access expression is substituted by symbolic value of the field.
For each field-access expression $v.f_i$ in the conjunct (line 5), if the current path condition implies $v$ is $\nil$,
the path condition is unsatisfiable and is discarded (lines 6-7).
In case the path condition implies $v$ is constrained by a points-to predicate,
it substitutes $v.f_i$ with the corresponding symbolic name for the field in the predicate (lines 8-9).
Otherwise, if $v$ is constrained by an inductive predicate, it unfolds the predicate to
find points-to predicate for $v$ (lines 10-14).
In the last case (lines 15-16), it considers 
the current path condition does not have enough
information to resolve $v.f_i$ and simply returns empty.
For field-assign expression $v.f_i := e$, after transforming the expression with above steps,
it substitutes
 the left-hand side with a fresh symbolic name $f_i'$,
update the mapping from $v.f_i$ (or $x.f_i$ in case 
$\D{\implies} x = v$)
to $f_i'$, then change $:=$ to $=$ (lines 17-20).
Note that the update at line 19 may override the update at line 9 for left-hand side.
Similar to Algorithm~\ref{algo.genSpec}, the correctness of Algorithm~\ref{algo.field.acc} follows
from the fact that each final path condition is an under-approximation of the original path condition
because of the unfolding process.
\input{sections/algo-field-acc.tex}
For instance, the path condition
$\seppred{\code{bst}}{this\_root,minE,maxE} \wedge t=this\_root \wedge t\neq\nil \wedge x<t.element$
has field-access expression $t.element$
which need to be transformed. Using Algorithm~\ref{algo.field.acc}, we get the final path condition which can be passed to the solver:
\[
\small
\begin{array}{l}
$$\exists elt, l, r.~\sepnodeF{this\_root}{BinaryNode}{elt, l, r} \sep \seppred{\code{bst}}{l, minE, elt} \sep \seppred{\code{bst}}{r, elt, maxE}~\wedge\\
\qquad minE < elt \wedge maxE > elt \wedge t = this\_root \wedge t \neq \nil \wedge x < elt$$
\end{array}
\]
The solver verifies that the path condition is satisfiable and then returns a model which is a $BinarySearchTree$ with 1 node.
The $element$ field of the node has value $1$ and the value of parameter $x$ is 0.
\repconf{The details of the transformation are shown in App. 3.}{}



%% file: sections/prog-fig.tex
\begin{figure}[t]
\centering
\small
\[
\begin{array}{rcl}
\myit{datat} & ::= & \btt{data} ~c ~\{~ {(type~v;)}^* ~\} \\
 \myit{type} & ::= & 
 \bool ~|~ \int ~|~ c \\
\form{\myit{prog} & ::= & stmt^*}  \\
stmt & ::= &  v~{:=}~e ~|~  v{.}f_i~{:=}~e ~|~ \code{goto}~ e
~|~ \code{assert} ~e ~|~ \myif{e_0}{\code{goto}~ e_1}{\code{goto}~e_2} \\
& & |~v := \btt{new}~c(v_1, ..., v_n)~|~\btt{free}~v
  \\ 
e & ::= &  k~|~ v ~|~ v{.}f_i ~|~ e_1 ~op_{b}~ e_2 ~|~
op_{u}~e ~|~\nil
\end{array}
\]
\caption{A core intermediate language}\label{fig.syntax}
\end{figure}

%% file: sections/rules.tex
\begin{figure}[t]
{\small \begin{center}
{\scriptsize
\begin{minipage}{0.98\textwidth}
\begin{frameit} 
\[
\hlr{C-CONST}{
}
{\evalse{{\sstack}}{k}{k}}
\quad
\hlr{C-VAR}{
}
{\evalse{{\sstack}}{v}{k}}
\quad
\hlr{C-NULL}{
}
{\evalse{{\sstack}}{\nil}{\nil}}
\]
\[
\hlr{C-UNOP}{
\evalse{\sstack}{e}{k} \quad
}
{\evalse{\sstack}{op_{u}~ e}{ ~op_{u}~ k}}
\quad
\hlr{C-BINOP}{
\evalse{{\sstack}}{e_1}{k_1} \quad
\evalse{{\sstack}}{e_2}{k_2} \quad
}
{\evalse{{\sstack}}{e_1 ~op_{b}~ e_2}{k_1 ~op_{b}~ k_2 }}
\]
\[
\hlr{C-LOAD}{
\begin{array}{c}
\evalse{{\sstack}}{v}{l} \quad
\evalse{{\sstack}}{l.f_i}{k}
\end{array}
}
{\evalse{{\sstack}}{v.f_i}{k}}
\quad
 \hlr{C-FREE}{
\evalse{{\sstack}}{v}{l}
\quad
\sstack' {=} \sstack \setminus \{l.f_i \mapsto \_\} ~\forall i {=} 1..n
\quad
\iota = \Sigma(pc+1)
}
{\hquad{\sectx{\Sigma}{\sstack}{\D}{pc}{\code{true}}{\btt{free}~v}}{\sectx{\Sigma}{\sstack'}{\D}{pc+1}{\code{true}}{\iota}}}
\]
\[
\hlr{C-ASSIGN}{
\evalse{{\sstack}}{e}{k} ~~~
\sstack' {=} \sstack[v \leftarrow k]
~~~
\text{fresh } v' ~~~
\form{e'} {=} \form{e}[v'/v] ~~~
\D' {\equiv} \exists v'. \D[v'/v] \wedge v{=}\form{e'}
~~~
\iota {=} \Sigma[pc+1]
}
{\hquad{\sectx{\Sigma}{\sstack}{\D}{pc}{\code{true}}{v := e}}
{\sectx{\Sigma}{\sstack'}{\D'}{pc+1}{\code{true}}{\iota}}}
\]
\[
\hlr{C-NEW}{
\begin{array}{c}
\text{fresh } l \quad \text{fresh } v' \quad
\D' {\equiv} \exists v'. \D\subst{v'}{v} \sep \sepnodeF{v}{c}{v_1,...,v_n} \\
\sstack'_1 {=} \sstack[l.f_i \leftarrow ({\sstack}\vdash{v_i})]~\forall i {=} 1..n \quad
\sstack' {=} \sstack'_1[v \leftarrow l]
\quad \iota {=} \Sigma(pc{+}1)
\end{array}
}
{\hquad{\sectx{\Sigma}{\sstack}{\D}{pc}{\code{true}}{v~ {=}~\code{new}~c(v_1,...,v_n)}}{\sectx{\Sigma}{\sstack'}{\D'}{pc{+}1}{\code{true}}{\iota}}}
\]
\[
\hlr{C-STORE}{
\begin{array}{c}
\evalse{{\sstack}}{v}{l} \quad \evalse{{\sstack}}{e}{k} \quad
\sstack' {=} \sstack[l.f_i \leftarrow k]
\quad
\D' {\equiv} \D \wedge v.f_i {:=} e \quad
\iota{=}\Sigma(pc+1) \\
\end{array}
}
{\hquad{\sectx{\Sigma}{\sstack}{\D }{pc}{\code{true}}{v.f_i = e}}{\sectx{\Sigma}{\sstack'}{\D'}{pc+1}{\code{true}}{\iota}}}
\]
\[
\hlr{C-GOTO}{
\evalse{\sstack}{e}{k} \quad
\iota{=}\Sigma(k)
}
{\hquad{\sectx{\Sigma}{\sstack}{\D}{pc}{\code{true}}{\code{goto} ~e}}
{\sectx{\Sigma}{\sstack}{\D}{k}{\code{true}}{\iota}}}
\]
\[
\hlr{C-ASSERT}{
\evalse{{\sstack}}{e}{\code{true}} \quad \D' {\equiv} {\D}\wedge e
\quad \iota{=}\Sigma(pc{+}1)
}
{\hquad{\sectx{\Sigma}{\sstack}{\D}{pc}{\code{true}}{\code{assert}(e)}}{\sectx{\Sigma}{\sstack}{\D'}{pc{+}1}{\code{true}}{\iota}}}
\]
\[
\hlr{C-TCOND}{
\evalse{\sstack}{e_0}{\code{true}} ~~
\evalse{\sstack}{e_1}{k_1} ~~
\evalse{\sstack}{e_2}{k_2} ~~
\D_1{\equiv}\D \wedge e_0 ~~
\D_2{\equiv}\D \wedge\neg e_0 ~~
\iota_1{=}\Sigma(k_1) ~~
\iota_2{=}\Sigma(k_2)
}
{
\begin{array}{l}
{\hquad{\sectx{\Sigma}{\sstack}{\D}{pc}{\code{true}}{\myif{e_0}{\code{goto} ~e_1}{\code{goto} ~e_2}}}{}}\\
{\quad\quad \sectx{\Sigma}{\sstack}{\D_1}{k_1}{\code{true}}{\iota_1},\sectx{\Sigma}{\sstack}{\D_2}{k_2}{\code{false}}{\iota_2}}
\end{array}
}
\]
\[
\hlr{C-FCOND}{
\evalse{{\sstack}}{e_0}{\code{false}} ~~
\evalse{\sstack}{e_1}{k_1} ~~
\evalse{\sstack}{e_2}{k_2} ~~
\D_1{\equiv}\D \wedge e_0 ~~
\D_2{\equiv}\D \wedge\neg e_0 ~~
\iota_1{=}\Sigma(k_1) ~~
\iota_2{=}\Sigma(k_2)
}
{
\begin{array}{l}
{\hquad{\sectx{\Sigma}{\sstack}{\D}{pc}{\code{true}}{\myif{e_0}{\code{goto} ~e_1}{\code{goto} ~e_2}}}{}}\\
{\quad\quad \sectx{\Sigma}{\sstack}{\D_1}{k_1}{\code{false}}{\iota_1},\sectx{\Sigma}{\sstack}{\D_2}{k_2}{\code{true}}{\iota_2}}
\end{array}
}
\]
\end{frameit}
\end{minipage}
}
\end{center}}
\caption{Execution rules: \form{\Sigma[x \leftarrow k]} updates the mapping \form{\Sigma} by setting \form{x} to be \form{k}; \form{\code{fresh}} is used as an overloading function
to return a new variable/address; 
\form{\evalse{\sstack}{e}{ k}}
denotes the evaluation of expression \form{e} to a concrete value \form{k}
in the current context \form{\sstack}}\label{fig.rules}
\end{figure}

%% file: sections/tree.tex
\begin{figure*}[t]
\centering
  \includegraphics[width=.15\linewidth]{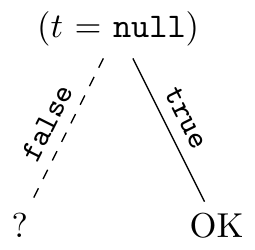}~~~~~~
  \includegraphics[width=.3\linewidth]{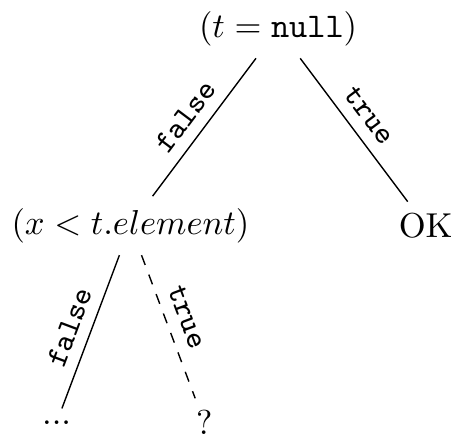}~~~~~~
  \includegraphics[width=.3\linewidth]{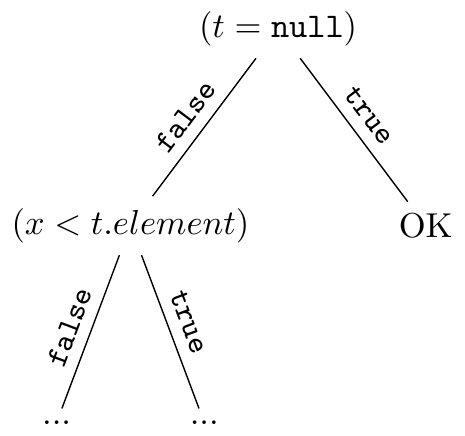} \\
  (a)~~~~~~~~~~~~~~~~~~~~~~~~~~~~~~~~~~~~~~~~~~~~~(b)~~~~~~~~~~~~~~~~~~~~~~~~~~~~~~~~~~~~~~~~~~~(c)
\caption{Constraint trees construction: a question mark represents
an unexplored path and OK denotes the execution terminates without error}
\label{fig:tree}
\end{figure*}

%% file: sections/algo-field-acc.tex
\begin{algorithm}[t]
	\small

$map \leftarrow \emptyset$\\
\ForEach {$\sepnode{v}{c}{v_1,...,v_n} \in \D$} {
	$map \leftarrow map \cup \{v.f_i \leftarrow v_i\}$
}
\ForEach{$(lhs~op~rhs) \in \D$} {
	\ForEach{ $v.f_i\in (lhs~op~rhs)$} {
		\If{$\D {\implies} v = \nil$}{
			\Return $\emptyset$
		}
		\ElseIf{$map(v.f_i) = v_i~\|~map(x.f_i) = v_i~\&\&~\D{\implies} v=x$} {
			$(lhs~op~rhs) \leftarrow (lhs~op~rhs) \subst{v_i}{v.f_i}$
		}
		\ElseIf{$\code{P}(\setvars{v}) \in \D ~\&\&~ (v\in\setvars{v}~\|~x\in\setvars{v} ~\&\&~ \D{\implies} v=x)$} {
			$\D_s \leftarrow \code{unfold}(\D,\code{P}(\setvars{v}))$, $\Gamma \leftarrow \emptyset$\\
			\ForEach{$\D_i \in \D_s$} {
				$\Gamma \leftarrow \Gamma \cup \code{preprocess}(\D_i)$
			}
			\Return{$\Gamma$}
		}
		\Else {
			\Return $\emptyset$
		}
	}
	\If{$op ~is~ :=$} {
		Substitute $lhs$ with a fresh symbolic name\\
		Update the field in $map$ to the new name\\
		Substitute $:=$ with $=$
	}
}
\Return{$\{\D\}$}
\caption{$\code{preprocess}(\Delta)$} \label{algo.field.acc}
\end{algorithm}

%% file: sections/experiment.tex
We have implemented our proposal in a tool, named Concolic StarFinder (CSF), 
with 6770 lines of Java code 
as a module inside the Java PathFinder framework. In the following, we conduct three experiments 
and contrast CSF's performance with existing approaches. All experiments are conducted on a laptop with 2.20GHz 
 and 16 GB RAM.
\repconf{Instructions to obtain the artifact, which contains the tool source code, benchmarks and test scripts to replicate the experiments, are included in App. 4.\\}{\\}

\noindent \textbf{First Experiment} In this experiment, we assume CSF is used as a stand-alone tool to generate test inputs for heap-manipulating programs. That is, the users provide a program and a precondition,
then apply CSF to automatically generate a set of test inputs. The experimental subject is a comprehensive set of benchmark programs collected from previous publications,
which includes \emph{Singly-Linked List} (SLL), \emph{Doubly-Linked List} (DLL), \emph{Stack}, \emph{Binary Search Tree} (BST), \emph{Red Black Tree} (RBT) from SIR~\cite{sir}, \emph{AVL Tree}, \emph{AA Tree} (AAT) from Sireum/Kiasan~\cite{sireum},
\emph{Tll} from~\cite{Le:CAV:2014}, 
the motivation example from SUSHI~\cite{Braione:ISSTA:2017},
the TSAFE project~\cite{tsafe}, and the Gantt project~\cite{gantt}. 
In total,
we have 74 methods whose line of codes range from dozens to more than one thousand. 
For each method, the precondition according to the original publication is adopted for generating test inputs using CSF.
In the specification-based testing stage, CSF is configured to generate all test inputs
with a depth of 1 (e.g., unfolding the precondition once). 

We compare CSF with two state-of-the-art tools, e.g.,
JBSE~\cite{Braione:2016:JSE:2950290.2983940} and BBE~\cite{Visser:ISSTA:2004}.
JBSE uses HEX for specifying the invariants of valid test inputs and generates test inputs accordingly. We use the same invariants reported in~\cite{Braione:2016:JSE:2950290.2983940} in our experiments. Note that because the HEX invariants for \emph{SLL}, \emph{Stack}, \emph{BST}, \emph{AA Tree} and \emph{Tll} are not available\footnote{and it is unclear to us whether HEX is capable to specify them.}, we skip running JBSE with these test subjects. BBE is explained in Section~\ref{sec.mov.ex}. In the following, we answer multiple research questions (RQ) through experiments. \\


\noindent \emph{RQ1: Does CSF generate valid test inputs?}
We apply CSF to generate test inputs for the 74 methods. To check whether the generated test inputs are valid, we validate the generated test inputs with the $repOK$ method in the data structures. The results are shown in the columns named $\#Tests$ in Table~\ref{effectiveness} for each test subject.
The entries for JBSE and BBE are in the form of the number of valid test inputs over the total number of test inputs.
As expected, all test inputs generated by CSF are valid. In comparison, JBSE generates 4.65\% valid test inputs and BBE generates 7.83\% valid test inputs. The reason for the poor results of JBSE and BBE is that the reference variables/fields are
initialized with the wrong values
or never initialized if they are not accessed. Note that by default, JBSE generates partially initialized test inputs,
so we additionally call method $repOK$ to concretize them.
CSF solves the path conditions, which contain the precondition, to generate test inputs, which are guaranteed to be valid. We thus conclude that using an expressiveness language is important in achieving validity. \\

\input{sections/tableFull.tex}

\noindent \emph{RQ2: Can CSF achieve high code coverage?} We use JaCoCo~\cite{jacoco} to measure the branch coverage of the generated test inputs.
The results are shown in the sub-columns named $Cov.(\%)$ (which is the coverage achieved by valid test inputs) and $NCov.(\%)$
(which is the coverage achieved by all test inputs including the invalid ones) in Table~\ref{effectiveness}. The winners are highlighted in bold. Note that for CSF, because all the test inputs are valid, we omit the
column $NCov.(\%)$. The results show that CSF achieves nearly 100\% branch coverage for almost all programs except TSAFE, whose coverage is 59.46\%.
For 70 out of 74 methods, CSF can obtain 100\% branch coverage (including
branches for auxiliary methods and excluding infeasible branches). CSF fails to cover 1 branch in two methods (i.e., $remove$
for $RBT$ and $remove$ for $AAT$) and 3 branches in one method (i.e., $put$ for $RBT$). The reason is that although the path conditions
leading to those branches are satisfiable, the solver times out. For method $TS\_R\_3$, CSF achieves 59.46\% branch coverage because
in the execution, some native methods are invoked and applying symbolic execution to those paths are infeasible. Moreover, some of the path conditions contain string constraints which are not supported by the solver.
For JBSE and BBE, the average coverage is 68.54\% and
37.85\% respectively if we consider valid test inputs only. If all test inputs are considered, the average coverage increases to 95.59\%
for JBSE and 54.66\% for BBE. Note that the coverage is inflated with invalid test inputs. \\ 

\noindent \emph{RQ3: Is CSF sufficiently efficient?} We measure the time needed to generate test inputs (sub-columns $T(s)$ in the Table~\ref{effectiveness}).
The results show that CSF needs 57.34 seconds on average for each program. 
The numbers for JBSE and BBE are 8.75 and 9.50
seconds respectively.
Both JBSE and BBE are faster than CSF since they solve simpler constraints (e.g., without inductive predicates). However, their efficiency has a cost in term of the validity of the generated test inputs and the achieved code coverage. To conclude, we believe CSF is sufficiently efficient to be used in practice. We further show the number
of solver calls used in CSF, i.e., the sub-column $\#Calls$ in the Table~\ref{effectiveness}. 
The results are represented in form of the number of solver calls for specification-based testing over that of concolic execution.
The results show that CSF needs 43 calls in average. Note that the number of solver calls in the specification-based testing stage varies according to the number of disjuncts
in the precondition. \\

\noindent \textbf{Second Experiment} One infamous limitation of symbolic execution testing approach is it cannot handle programs with complex numerical conditions. On the other hand, specification-based testing approach does not suffer this limitation because it generates test inputs independently of programs under test. In this experiment, we aim to show the usefulness of specification-based testing in CSF, especially for programs with complex numerical conditions. To do that, we systematically compose a set of programs which travel a singly-linked list, apply a method from \emph{java.lang.Math} library to the list elements, and check if the result satisfies some condition. One example is shown in Fig.~\ref{random} with method \emph{cos}, which returns the cosin value of an integer. In total, we have 32 programs with 32 different methods from \emph{java.lang.Math} library. We run CSF with only specification-based testing (to generate 10 test inputs) and compare the results with BBE. We cannot compare with JBSE because we do not have the HEX invariant for singly-linked list. However, we note that JBSE is a symbolic execution engine, which means it has difficulties in handling complex numerical conditions. The list elements has random values from -32 to 31 for all the tools. Due to randomness, we repeat the experiment 10 times for each program.

\begin{figure}[t]
\centering
\begin{minipage}[h]{0.85\linewidth}
\begin{lstlisting}[numbers=none]
public boolean withCos(Node root) {
	while (root != null) {
		if (Math.cos(root.elem) == 1) return true;
		root = root.next; }
	return false;
}
\end{lstlisting}
\end{minipage}
\caption{An example in the second experiment}
\label{random}
\end{figure}

In average, while CSF obtains 88.28\% branch coverage, BBE obtains 75.31\%. 
 The average number of solver calls is 18 and the average time is 2.27 seconds for each program. 
For BBE, it generates 10 test inputs for each program but only 4 of them satisfy \emph{repOK} in 2.97 seconds. 
From the results, we conclude that the specification-based testing phase is useful, especially for programs with complex numerical conditions. \\

\noindent \textbf{Third Experiment}
Although having a specification language based on separation logic allows us to precisely specify preconditions of the programs under test and generate valid test inputs,
it could be non-trivial for ordinary users to use such a language. This problem has been recognized by the community and there have been multiple approaches to solve this problem~\cite{infer,Le:CAV:2014,DBLP:conf/icsm/LeLLG16,Fragoso:POPL19}. One noticeable example which has made industrial impact is the Infer static analyzer~\cite{infer}, which infers preconditions of programs through 
bi-abduction~\cite{Calcagno:JACM:2011}. In this experiment, we show that CSF can be effectively combined with Infer so that CSF can be applied without user-specified preconditions.

We first apply Infer to generate preconditions of the programs under test and then apply CSF to generate test inputs accordingly.
%
The test subject is PLEXIL~\cite{plexil}, i.e., NASA's plan automation and execution framework. Specifically, we
analyze its verification environment PLEXIL5~\cite{plexil5} with Infer, and collect 88 methods that 
have explicit preconditions returned by Infer. 

The experimental results are shown in Table~\ref{tbl:plexil}, which are categorized based on the number of initial test inputs generated
from Infer's preconditions (column $\#Init~Tests$). The second column $\#Methods$ shows the number of methods in the category.
The column $\#Tests$ shows the number of generated test inputs and the column $\#Exceptions$ shows the number of exceptions in the category.
Lastly, two columns
$\#Calls$ and $Time(s)$ show the number of solver calls and the time needed to generate the test inputs respectively.
In summary, CSF generates 292 test inputs in 344 seconds which achieved 58.36\% branch coverage in average.
Our investigation shows that all of these test inputs are valid according to the inferred preconditions.
Interestingly, 261 out of the 292 test inputs (i.e., 89\%) lead to $RuntimeException$ during execution.
The interpretation can be either (1) the inferred preconditions are too weak
to capture all the necessary conditions for valid test inputs generation, or (2) there are potential bugs in the programs.

\begin{table}[t]
		\small
    \centering
	\caption{Experiment 3 with Infer: Results}
        \begin{tabular}{| c | c | c | c | c | c |}
			\hline
			\#Init Tests & \#Methods & \#Tests & \#Exceptions & \#Calls & Time(s) \\
			\hline
    1 & 8 & 10 & 10 & 8/14 & 16 \\
    2 & 51 & 130 & 119 & 102/206 & 167 \\
	3 & 29 & 152 & 132 & 87/254 & 161 \\
	\hline
    \end{tabular}
\label{tbl:plexil}
\end{table}

To give an example, method $integerValue$ receives an Abstract Syntax Tree (AST) as input and the AST must contain an \emph{INT} token. The inferred precondition only says that the input should not be $\nil$. One of the test inputs generated by CSF is shown in Fig.~\ref{plexil}. The execution result is $RuntimeException$ because the value of field $ttype$ does not match with the value of \emph{INT} token, which is 108.

 It would be interesting to develop a full integration of CSF and 
 the recent bi-abduction for erroneous specification inference
\cite{Fragoso:POPL19} so that we can generate meaningful test inputs automatically
to witness bugs for any program.

\begin{figure}[t]
\centering
\begin{minipage}[h]{0.85\linewidth}
\begin{lstlisting}[numbers=none]
public void test_integerValue1() throws Exception {
	PlexilTreeParser obj = new PlexilTreeParser();
	plexil.PlexilASTNode _t = new plexil.PlexilASTNode();
	obj.ASTNULL = new antlr.ASTNULLType();
	int ttype_1 = 0;
	plexil.PlexilASTNode right_3 = null;
	plexil.PlexilASTNode down_2 = null;
	_t.ttype = ttype_1; _t.down = down_2; _t.right = right_3;
	obj.integerValue(_t);
}
\end{lstlisting}
\end{minipage}
\caption{A test input which leads to $RuntimeException$}
\label{plexil}
\end{figure}


%% file: sections/tableFull.tex
\begin{table}[t]
	\scriptsize
    \centering
	\caption{Experiment 1 \& 2: Results}
	        \begin{tabular}{| l | c | c | c | c | H H H H c | c | c | c | c | c | c | c |}
				\hline
	    \multirow{2}{*}{Program} & \multicolumn{4}{c|}{CSF} & \multicolumn{4}{H}{JSF} & \multicolumn{4}{c|}{JBSE} & \multicolumn{4}{c|}{BBE} \\
				\cline{2-17}
	             & \#Tests & Cov.(\%) & \#Calls & T(s) & \#Tests & Cov.(\%) & \#Calls & T(s) & \#Tests & Cov.(\%) & NCov.(\%) & T(s) & \#Tests & Cov.(\%) & NCov.(\%) & T(s) \\
    \hline
    DLL & 75 & \textbf{100} & 40/58 & 32 & 74 & \textbf{100} & 325 & 49 & 121/5146 & 56 & 100 & 206 & 0/35 & 0 & 21 & 21 \\
    \hline
    AVL & 62 & \textbf{100} & 36/654 & 274 & 69 & \textbf{100} & 623 & 400 & 76/295 & \textbf{100} & 100 & 48 & 17/117 & 70 & 89 & 69 \\
    \hline
    RBT & 133 & \textbf{99} & 14/1106 & 2403 & 314 & \textbf{100} & 2070 & 2256 & 137/291 & 87 & 91 & 38 & 14/380 & 26 & 53 & 333 \\
    \hline
    SUSHI & 5 & \textbf{100} & 3/38 & 8 & 7 & \textbf{100} & 30 & 5 & 0/900 & 0 & 100 & 24 & 2/27 & 25 & 25 & 8 \\
    \hline
    TSAFE & 16 & \textbf{59} & 1/595 & 1190 & 5 & 24 & 13 & 3 & 0/32 & 0 & 5 & 10 & 0/1 & 0 & 0 & 1 \\
    \hline
    Gantt & 22 & \textbf{100} & 2/156 & 25 & 21 & \textbf{100} & 140 & 25 & 17/887 & 55 & 90 & 24 & 0/6 & 0 & 5 & 2 \\
    \hline
    SLL & 29 & \textbf{100} & 21/8 & 11 & 26 & \textbf{100} & 55 & 11 & - & - & - & - & 16/50 & 66 & 71 & 19 \\
    \hline
    Stack & 18 & \textbf{100} & 16/2 & 7 & 18 & \textbf{100} & 31 & 7 & - & - & - & - & 11/14 & 84 & 84 & 6 \\
    \hline
    BST & 47 & \textbf{100} & 16/33 & 14 & 182 & \textbf{100} & 698 & 241 & - & - & - & - & 19/260 & 69 & 86 & 131 \\
    \hline
    AAT & 46 & \textbf{99} & 21/352 & 277 & 103 & \textbf{100} & 1179 & 1981 & - & - & - & - & 3/166 & 6 & 43 & 111 \\
    \hline
    Tll & 6 & \textbf{100} & 2/4 & 2 & 3 & \textbf{100} & 11 & 2 & - & - & - & - & 1/4 & 38 & 50 & 2 \\
    \hline
    Math & 320 & \textbf{88} & 576/0 & 73 & 78 & 78 & 196 & 18 & - & - & - & - & 128/320 & 75 & 79 & 95 \\
    \hline
    \end{tabular}
\label{effectiveness}
\end{table}

%% file: sections/relatedWork.tex
We review  closely related work in the following, emphasis is given to approaches that generate test inputs for heap-manipulating programs. \\

\noindent \emph{Concolic testing programs with heap inputs}
This work is the first work that uses separation logic for concolic testing.
The engineering design of our tool is based on that of JDart~\cite{tacas2016-ldghikrr}. However, JDart, like most concolic execution engines, 
e.g.,~\cite{Godefroid:2005:DDA:1065010.1065036,Godefroid:2012:SWF:2090147.2094081,DBLP:conf/nfm/JayaramanHGK09,Marinescu:2012:MTS:2337223.2337308,Tanno:2015:TCA:2819009.2819147},
does not support data structures as symbolic input for testing methods. 
Our work is related to CUTE~\cite{Sen2005} and Pex~\cite{Tillmann:2008:PWB:1792786.1792798}. CUTE~\cite{Sen2005} does support data structures as input by using the so-called \emph{logical input map} to keep track of input memory graph.
However, CUTE cannot handle unbounded inputs
nor capture the shape relations between pointers,
which leads to imprecision.
Pex~\cite{Tillmann:2008:PWB:1792786.1792798} uses a type system~\cite{Vanoverberghe:TACAS:2009} to describe disjointness of memory regions.
But again, Pex cannot handle unbounded inputs. Moreover, the type system can only reason about the \emph{global} heap, which leads to complex constraints and hence poor scalability. 
In comparison, our work handles unbounded inputs and shape relations are well-captured by separation logic predicates.\\


\noindent \emph{Lazy initialization}
As far as we know, {\lazy}~\cite{Khurshid2003} is the only way to handle unbounded inputs.
However, most works in this direction, e.g.,~\cite{Deng:ASE:2006,Deng:2007:TCS:1306879.1307404,Hillery:2016:EHS:2963187.2963200,Visser:ISSTA:2004}, did not address the problem of generating invalid test inputs due to the lack of constraints on the shapes
of the input data structures.
This work is related to the tool JSF presented in 
\cite{DBLP:journals/corr/Pham17,Pham:2018:THP:3183440.3194964}. 
 While 
JSF uses separation logic for specifying preconditions and apply
classical symbolic execution,
 ours relies on concolic execution. Moreover, to support memory access, JSF unfolds those heaps accessed by reference variables in advance, our work prepares heap accesses via lazy unfolding 
which
helps to encode both executed/not-yet-executed paths and heap accesses together. 
Another related work is~\cite{Braione:2015:SEP} by Braione \emph{et al.}, which we have discussed extensively in previous sections. The logic presented in~\cite{Braione:2015:SEP}, HEX, is not expressive enough to describe many popular data structures, including the binary search tree in our motivating example. \\

\noindent \emph{Specification-based testing}
has been an active research area for decades. Depending on the testing goals, different types of logic have been used as the specification languages to generate test inputs, for example Alloy~\cite{Marinov:ASE:2001}, Java predicates~\cite{Boyapati:2002:KAT:566172.566191}, and temporal logic~\cite{Heimdahl:2004,Hong:TACAS:2002}. However, we are not aware of any existing work that generate test inputs from the specification 
in separation logic like ours.\\

\noindent \emph{Separation logic}
Research in separation logic focuses on
static verification~\cite{Calcagno:JACM:2011,Chin:SCP:2012,Le:CAV:2014,DBLP:conf/tacas/Le0Q18,Piskac:2013:ASL:2526861.2526927}, which may return false positives and are not able to generate test inputs. 

%% file: sections/conclusion.tex
We have presented a novel concolic execution engine for heap-manipulating programs
based on separation logic. Our engine starts with generating a set of initial test inputs based on preconditions.
It concretely executes, monitors the executions
and generates new inputs to drive the execution to unexplored code.
We have implemented the proposal in {\tool} and evaluated it over benchmark programs.
The experimental results
show {\tool}'s effectiveness and practical applications.\\


%% file: sections/tounit.tex
In the following, we show how to transform the symbolic model into a test input using procedure {\tounit}.
In this  transformation,
we
maintain a list of initialized variables.
The transformation has three steps.
%
 Firstly, for each points-to predicate \form{\sepnodeF{v}{c}{...}},
we create a new object of type $c$ and assign the new object to $v$. Similarly,
for each predicate $v=\nil$ or $v=k$,
 we assign $\nil$ or $k$ to $v$
respectively. After that, we add $v$ into the list of initialized variables.
%
 Secondly, for each equality predicate $v_1=v_2$,
in case 
either $v_1$ or $v_2$ is not initialized, we find an
initialized alias $v$ for $v_1$ and $v_2$ in the model, then assign $v$ to
$v_1$ and $v_2$. In case $v_1$ and $v_2$ are not alias with any
initialized variable, we create a new object with compatible type and assign it to $v_1$ and $v_2$.
After that, both $v_1$ and $v_2$ are added into the list of initialized variables.
%
 Lastly, for each points-to predicate \form{\sepnodeF{v}{c}{v_1,...,v_n}},
we assign $v_i$ to $v.f_i$ for $i = 1..n$. Note that before this step,
all variables $v$, $v_1$, ..., $v_n$ are already initialized given the previous two steps.

%% file: sections/appC.tex
We show the details of tranformation for the path condition
\[
\small
\begin{array}{l}
\seppred{\code{bst}}{this\_root,minE,maxE} \wedge t=this\_root \wedge t\neq\nil \wedge x<t.element
\end{array}
\]
with field-access expression $t.element$.

From the path condition, we know that $t$ is alias with $this\_root$ and is constrained by the predicate
$\seppred{\code{bst}}{this\_root,minE,maxE}$, so we unfold the predicate and get two new path conditions:
\[
\small
\begin{array}{ll}
1. & \emp \wedge this\_root = \nil \wedge t = this\_root \wedge t \neq \nil \wedge x < t.element \\
2. & \exists elt, l, r.~\sepnodeF{this\_root}{BinaryNode}{elt, l, r} \sep \seppred{\code{bst}}{l, minE, elt} \sep \seppred{\code{bst}}{r, elt, maxE}~\wedge \\
& \qquad minE < elt \wedge maxE > elt \wedge t = this\_root \wedge t \neq \nil \wedge x < t.element
\end{array}
\]
In the first case, $this\_root = \nil \wedge t = this\_root$ so we cannot have symbolic value for $t.element$
and get rid of this path condition. Note that in this case, the path condition is unsatisfiable because
it also contains $t \neq \nil$.
In the second case, $t$ is alias with $this\_root$, which points to a $BinaryNode$ with the symbolic value for field $element$ is $elt$,
so we substitute $t.element$ with $elt$ and get the final path condition which can be passed to the solver:
\[
\small
\begin{array}{l}
$$\exists elt, l, r.~\sepnodeF{this\_root}{BinaryNode}{elt, l, r} \sep \seppred{\code{bst}}{l, minE, elt} \sep \seppred{\code{bst}}{r, elt, maxE}~\wedge \\
\qquad minE < elt \wedge maxE > elt \wedge t = this\_root \wedge t \neq \nil \wedge x < elt$$
\end{array}
\]

%% file: sections/artifact.tex
We provide a Docker image containing the tool source code, benchmarks and test scripts to replicate our experiments. Instructions to install Docker on various platforms can be found in this link:
\url{https://docs.docker.com/install/}.\\

\noindent Depending on the way Docker is installed, all the following Docker commands may need to be run with \textbf{sudo}.

\begin{enumerate}
\item Pulling the Docker image from Docker Hub\\
	\texttt{docker pull artifact2019/concolic}
\item Creating a Docker container\\
	\texttt{docker run -ti artifact2019/concolic /bin/bash}\\
    after this step, you will be inside the container and the current directory is\\ \texttt{/tools/jpf-costar}
\item Running all the examples from the current directory\\
	\texttt{bin/testAll.sh}
\item The generated test inputs will be in the directory\\
	\texttt{src/output}
\end{enumerate}

%% file: ms.bbl
\begin{thebibliography}{10}
\providecommand{\url}[1]{\texttt{#1}}
\providecommand{\urlprefix}{URL }
\providecommand{\doi}[1]{https://doi.org/#1}

\bibitem{PySymEmu}
{A fuzzer and a symbolic executor walk into a cloud}.
  \url{https://blog.trailofbits.com/2016/08/02/engineering-solutions-to-hard-program-analysis-problems/}

\bibitem{infer}
{Facebook Infer}. \url{https://fbinfer.com/}

\bibitem{gantt}
{GanttProject}. \url{https://github.com/bardsoftware/ganttproject}

\bibitem{jacoco}
{JaCoCo}. \url{https://www.eclemma.org/jacoco/}

\bibitem{plexil}
{PLEXIL}. \url{http://plexil.sourceforge.net}

\bibitem{plexil5}
{PLEXIL5}. \url{https://github.com/nasa/PLEXIL5}

\bibitem{sir}
{SIR}. \url{http://sir.unl.edu/portal/index.php}

\bibitem{sireum}
{Sireum}. \url{https://code.google.com/archive/p/sireum/downloads}

\bibitem{Boyapati:2002:KAT:566172.566191}
Boyapati, C., Khurshid, S., Marinov, D.: {Korat: Automated Testing Based on
  Java Predicates}. In: Frankl, P.G. (ed.) ISSTA 2002, pp. 123--133. ACM
  (2002). \doi{10.1145/566172.566191}

\bibitem{Braione:ISSTA:2017}
Braione, P., Denaro, G., Mattavelli, A., Pezz\`{e}, M.: {Combining Symbolic
  Execution and Search-based Testing for Programs with Complex Heap Inputs}.
  In: Bultan, T., Sen, K. (eds.) ISSTA 2017, pp. 90--101. ACM (2017).
  \doi{10.1145/3092703.3092715}

\bibitem{Braione:2015:SEP}
Braione, P., Denaro, G., Pezz\`{e}, M.: {Symbolic Execution of Programs with
  Heap Inputs}. In: Nitto, E.D., Harman, M., Heymans, P. (eds.) FSE 2015, pp.
  602--613. ACM (2015). \doi{10.1145/2786805.2786842}

\bibitem{Braione:2016:JSE:2950290.2983940}
Braione, P., Denaro, G., Pezz\`{e}, M.: {JBSE: A Symbolic Executor for Java
  Programs with Complex Heap Inputs}. In: Zimmermann, T., Cleland{-}Huang, J.,
  Su, Z. (eds.) FSE 2016, pp. 1018--1022. ACM (2016).
  \doi{10.1145/2950290.2983940}

\bibitem{Calcagno:JACM:2011}
Calcagno, C., Distefano, D., O'Hearn, P.W., Yang, H.: {Compositional Shape
  Analysis by Means of Bi-Abduction}. JACM  \textbf{58}(6),  26:1--26:66
  (2011). \doi{10.1145/2049697.2049700}

\bibitem{Chin:SCP:2012}
Chin, W.N., David, C., Nguyen, H.H., Qin, S.: {Automated Verification of Shape,
  Size and Bag Properties via User-defined Predicates in Separation Logic}.
  Sci. Comput. Program.  \textbf{77}(9),  1006--1036 (2012).
  \doi{10.1016/j.scico.2010.07.004}

\bibitem{Deng:ASE:2006}
Deng, X., Lee, J., Robby: {Bogor/Kiasan: A K-bounded Symbolic Execution for
  Checking Strong Heap Properties of Open Systems}. In: ASE 2006, pp. 157--166.
  IEEE Computer Society (2006). \doi{10.1109/ASE.2006.26}

\bibitem{Deng:2007:TCS:1306879.1307404}
Deng, X., Robby, Hatcliff, J.: {Towards A Case-Optimal Symbolic Execution
  Algorithm for Analyzing Strong Properties of Object-Oriented Programs}. In:
  SEFM 2007. IEEE Computer Society (2007). \doi{10.1109/SEFM.2007.43}

\bibitem{tsafe}
Dennis, G.D.: {TSAFE : Building a Trusted Computing Base for Air Traffic
  Control Software}. Master's thesis, Massachusetts Institute of Technology,
  USA (2003)

\bibitem{Godefroid:2005:DDA:1065010.1065036}
Godefroid, P., Klarlund, N., Sen, K.: {DART: Directed Automated Random
  Testing}. In: Sarkar, V., Hall, M.W. (eds.) PLDI 2005, pp. 213--223. ACM
  (2005). \doi{10.1145/1065010.1065036}

\bibitem{Godefroid:2012:SWF:2090147.2094081}
Godefroid, P., Levin, M.Y., Molnar, D.: {SAGE: Whitebox Fuzzing for Security
  Testing}. Queue  \textbf{10}(1),  20:20--20:27 (2012).
  \doi{10.1145/2090147.2094081}

\bibitem{Heimdahl:2004}
Heimdahl, M.P.E., Rayadurgam, S., Visser, W., Devaraj, G., Gao, J.:
  {Auto-generating Test Sequences Using Model Checkers: A Case Study}. In:
  Petrenko, A., Ulrich, A. (eds.) FATES 2003, pp. 42--59. Springer (2003).
  \doi{10.1007/978-3-540-24617-6\_4}

\bibitem{Hillery:2016:EHS:2963187.2963200}
Hillery, B., Mercer, E., Rungta, N., Person, S.: {Exact Heap Summaries for
  Symbolic Execution}. In: Jobstmann, B., Leino, K.R.M. (eds.) VMCAI 2016, pp.
  206--225. Springer (2016). \doi{10.1007/978-3-662-49122-5\_10}

\bibitem{Hong:TACAS:2002}
Hong, H.S., Lee, I., Sokolsky, O., Ural, H.: {A Temporal Logic Based Theory of
  Test Coverage and Generation}. In: Katoen, J.P., Stevens, P. (eds.) TACAS
  2002, pp. 327--341. Springer (2002). \doi{10.1007/3-540-46002-0\_23}

\bibitem{Ishtiaq:2001:BAL:360204.375719}
Ishtiaq, S.S., O'Hearn, P.W.: {BI as an Assertion Language for Mutable Data
  Structures}. In: Hankin, C., Schmidt, D. (eds.) POPL 2001, pp. 14--26. ACM
  (2001). \doi{10.1145/360204.375719}

\bibitem{DBLP:conf/nfm/JayaramanHGK09}
Jayaraman, K., Harvison, D., Ganesh, V., Kiezun, A.: {jFuzz: A Concolic
  Whitebox Fuzzer for Java}. In: Denney, E., Giannakopoulou, D., Pasareanu,
  C.S. (eds.) NFM 2009, pp. 121--125 (2009)

\bibitem{Khurshid2003}
Khurshid, S., P\u{a}s\u{a}reanu, C.S., Visser, W.: {Generalized Symbolic
  Execution for Model Checking and Testing}. In: Garavel, H., Hatcliff, J.
  (eds.) TACAS 2003, pp. 553--568. Springer (2003).
  \doi{10.1007/3-540-36577-X\_40}

\bibitem{King:1976:SEP:360248.360252}
King, J.C.: {Symbolic Execution and Program Testing}. Commun. ACM
  \textbf{19}(7),  385--394 (1976). \doi{10.1145/360248.360252}

\bibitem{Le:CAV:2014}
Le, Q.L., Gherghina, C., Qin, S., Chin, W.N.: {Shape Analysis via Second-Order
  Bi-Abduction}. In: Biere, A., Bloem, R. (eds.) CAV 2014, pp. 52--68. Springer
  (2014). \doi{10.1007/978-3-319-08867-9\_4}

\bibitem{Le:CAV:2016}
Le, Q.L., Sun, J., Chin, W.N.: {Satisfiability Modulo Heap-Based Programs}. In:
  Chaudhuri, S., Farzan, A. (eds.) CAV 2016, pp. 382--404. Springer (2016).
  \doi{10.1007/978-3-319-41528-4\_21}

\bibitem{DBLP:conf/tacas/Le0Q18}
Le, Q.L., Sun, J., Qin, S.: {Frame Inference for Inductive Entailment Proofs in
  Separation Logic}. In: Beyer, D., Huisman, M. (eds.) TACAS 2018, pp. 41--60.
  Springer (2018). \doi{10.1007/978-3-319-89960-2\_3}

\bibitem{DBLP:conf/cav/LeT0C17}
Le, Q.L., Tatsuta, M., Sun, J., Chin, W.: {A Decidable Fragment in Separation
  Logic with Inductive Predicates and Arithmetic}. In: Majumdar, R., Kuncak, V.
  (eds.) CAV 2017, pp. 495--517. Springer (2017).
  \doi{10.1007/978-3-319-63390-9\_26}

\bibitem{DBLP:conf/icsm/LeLLG16}
Le, X.D., Le, Q.L., Lo, D., {Le Goues}, C.: {Enhancing Automated Program Repair
  with Deductive Verification}. In: ICSME 2016, pp. 428--432. {IEEE} Computer
  Society (2016). \doi{10.1109/ICSME.2016.66}

\bibitem{tacas2016-ldghikrr}
Luckow, K., Dimja\v{s}evi\'c, M., Giannakopoulou, D., Howar, F., Isberner, M.,
  Kahsai, T., Rakamari\'c, Z., Raman, V.: {JDart: A Dynamic Symbolic Analysis
  Framework}. In: Chechik, M., Raskin, J.F. (eds.) TACAS 2016, pp. 442--459.
  Springer (2016). \doi{10.1007/978-3-662-49674-9\_26}

\bibitem{Marinescu:2012:MTS:2337223.2337308}
Marinescu, P.D., Cadar, C.: {Make Test-zesti: A Symbolic Execution Solution for
  Improving Regression Testing}. In: Glinz, M., Murphy, G.C., Pezz{\`{e}}, M.
  (eds.) ICSE 2012, pp. 716--726. {IEEE} Computer Society (2012).
  \doi{10.1109/ICSE.2012.6227146}

\bibitem{Marinov:ASE:2001}
Marinov, D., Khurshid, S.: {TestEra: A Novel Framework for Automated Testing of
  Java Programs}. In: ASE 2001, pp.~22--. IEEE Computer Society (2001).
  \doi{10.1109/ASE.2001.989787}

\bibitem{DBLP:journals/corr/Pham17}
Pham, L.H., Le, Q.L., Phan, Q.S., Sun, J., Qin, S.: {Enhancing Symbolic
  Execution of Heap-based Programs with Separation~Logic for Test Input
  Generation}. In: ATVA 2019. To appear

\bibitem{Pham:2018:THP:3183440.3194964}
Pham, L.H., Le, Q.L., Phan, Q.S., Sun, J., Qin, S.: {Testing Heap-based
  Programs with Java StarFinder}. In: Chaudron, M., Crnkovic, I., Chechik, M.,
  Harman, M. (eds.) ICSE 2018, pp. 268--269. ACM (2018).
  \doi{10.1145/3183440.3194964}

\bibitem{Piskac:2013:ASL:2526861.2526927}
Piskac, R., Wies, T., Zufferey, D.: {Automating Separation Logic Using SMT}.
  In: Sharygina, N., Veith, H. (eds.) CAV 2013, pp. 773--789. Springer (2013).
  \doi{10.1007/978-3-642-39799-8\_54}

\bibitem{Reynolds:LICS02}
Reynolds, J.: {Separation Logic: A Logic for Shared Mutable Data Structures}.
  In: LICS 2002, pp. 55--74. {IEEE} Computer Society (2002).
  \doi{10.1109/LICS.2002.1029817}

\bibitem{Fragoso:POPL19}
Santos, J.F., Maksimović, P., Sampaio, G., Gardner, P.: {JaVerT 2.0:
  Compositional Symbolic Execution for JavaScript}. {PACMPL}
  \textbf{3}({POPL}),  66:1--66:31 (2019). \doi{10.1145/3290379}

\bibitem{Sen2005}
Sen, K., Marinov, D., Agha, G.: {CUTE: A Concolic Unit Testing Engine for C}.
  In: Wermelinger, M., Gall, H.C. (eds.) FSE 2005, pp. 263--272. ACM (2005).
  \doi{10.1145/1081706.1081750}

\bibitem{DBLP:conf/ndss/StephensGSDWCSK16}
Stephens, N., Grosen, J., Salls, C., Dutcher, A., Wang, R., Corbetta, J.,
  Shoshitaishvili, Y., Kruegel, C., Vigna, G.: {Driller: Augmenting Fuzzing
  Through Selective Symbolic Execution}. In: NDSS 2016. The Internet Society
  (2016)

\bibitem{Tanno:2015:TCA:2819009.2819147}
Tanno, H., Zhang, X., Hoshino, T., Sen, K.: {TesMa and CATG: Automated Test
  Generation Tools for Models of Enterprise Applications}. In: Bertolino, A.,
  Canfora, G., Elbaum, S.G. (eds.) ICSE 2015, pp. 717--720. {IEEE} Computer
  Society (2015). \doi{10.1109/ICSE.2015.231}

\bibitem{Tillmann:2008:PWB:1792786.1792798}
Tillmann, N., De~Halleux, J.: {Pex-White Box Test Generation for .NET}. In:
  Beckert, B., H{\"{a}}hnle, R. (eds.) TAP 2008, pp. 134--153. Springer (2008).
  \doi{10.1007/978-3-540-79124-9\_10}

\bibitem{Vanoverberghe:TACAS:2009}
Vanoverberghe, D., Tillmann, N., Piessens, F.: {Test Input Generation for
  Programs with Pointers}. In: Kowalewski, S., Philippou, A. (eds.) TACAS 2009,
  pp. 277--291. Springer (2009). \doi{10.1007/978-3-642-00768-2\_25}

\bibitem{Visser:ISSTA:2004}
Visser, W., P\v{a}s\v{a}reanu, C.S., Khurshid, S.: {Test Input Generation with
  Java PathFinder}. In: Avrunin, G.S., Rothermel, G. (eds.) ISSTA 2004, pp.
  97--107. ACM (2004). \doi{10.1145/1007512.1007526}

\bibitem{wang2018towards}
Wang, X., Sun, J., Chen, Z., Zhang, P., Wang, J., Lin, Y.: {Towards Optimal
  Concolic Testing}. In: Chaudron, M., Crnkovic, I., Chechik, M., Harman, M.
  (eds.) ICSE 2018, pp. 291--302 (2018). \doi{10.1145/3180155.3180177}

\bibitem{DBLP:conf/uss/Yun0XJK18}
Yun, I., Lee, S., Xu, M., Jang, Y., Kim, T.: {QSYM : A Practical Concolic
  Execution Engine Tailored for Hybrid Fuzzing}. In: Enck, W., Felt, A.P.
  (eds.) {USENIX} Security 2018, pp. 745--761. {USENIX} Association (2018)

\end{thebibliography}
